\documentclass{ws-procs975x65}
\def\beq{\begin{equation}}
\def\eeq{\end{equation}}
\begin{document}

\title{Generalized geometry \\ and non-symmetric gravity}

%\author{Andrew B. Author$^*$ and Charles D. Author}

%\address{University Department, University Name,\\
%City, State ZIP/Zone, Country\\
%$^*$E-mail: ab\_author@university.com\\
%www.university\_name.edu}

\author{Branislav Jur\v co} 

\address{Mathematical Institute, Faculty of Mathematics and Physics,
Charles University, \\
Prague, 186 75, Czech Republic\\
E-mail: jurco@karlin.mff.cuni.cz}

\author{Fech Scen Khoo$^{1,}$
\footnote{contributed to proceedings of the 14th Marcel Grossmann Meeting on General Relativity} 
and Peter Schupp$^2$}

\address{Department of Physics and Earth Sciences, Jacobs University,\\
Bremen, 28759, Germany\\
$^1$E-mail: f.khoo@jacobs-university.de\\
$^2$E-Mail: p.schupp@jacobs-university.de
%www.university\_name.edu
}

\author{Jan Vysok\'y } 

\address{Mathematical Sciences Institute, Australian National University,\\
Canberra, ACT, Australia\\
E-mail: jan.vysoky@anu.edu.au}

\begin{abstract}
Generalized geometry provides the framework 
for a systematic approach to non-symmetric metric gravity theory
and naturally leads to an Einstein-Kalb-Ramond gravity theory
with totally anti-symmetric contortion. 
The approach is related to
the study of the low-energy effective closed string gravity actions.
\end{abstract}


\bodymatter

%%%%%%%%%%%%%%%%% now a standard article style for the most part

\section{Motivation}

%World Scientific has changed its title capitalization style for article titles: capitalize the first word of the title and any other words that would normally be capitalized, just like an ordinary sentence (without a terminating period).

A proposal for a gravity theory with non-symmetric metric began with an idea of Einstein 
to unify gravity and electromagnetism (Refs.~\refcite{ES,UFT}).
In general relativity, the metric of the Riemannian manifold is a symmetric bilinear form. Interpreted as an invertible map from the tangent to the cotangent space, it is natural to allow also an anti-symmetric part. While the original hope of Einstein that the anti-symmetric part of a non-symmetric metric tensor may be directly related to  the electromagnetic force turned out to be incorrect, there is nevertheless phenomenological interest in non-symmetric gravity theories. 
%Inspired by string theory 
%which takes into account simultaneously a symmetric metric and an antisymmetric two-form Kalb-Ramond $B$-field, 
%a non-symmetric gravity theory was proposed by Damour et. al. (Ref.~\refcite{DDM}) in a similar spirit.
Damour {\it et al.} (Ref.~\refcite{DDM})
discussed the problems associated with the construction of non-symmetric gravity theories, 
where theories were typically in need for treatment of ghost terms.
There have since been numerous studies on the topic 
(Refs.~\refcite{Mof,JP,Nik}).
On the other hand, generalized geometry (Refs.~\refcite{Hitchin,Gua}),
which incorporates symmetries of string theory (T-duality, $B$-transform) and spacetime geometry (diffeomorphisms) seems to offer a well-suited geometrical framework for string theory as well as non-symmetric gravity theories.
Generalized geometry as an extension of Riemannian geometry can reproduce the Einstein-Hilbert and supergravity actions (Refs.~\refcite{Wal,Blu}).
In the present work, we consider an alternative approach that naturally incorporates torsion. For a recent work on metric connections with skew torsion  in Riemannian geometry, see Ref.~\refcite{Ana}. 
An alternative approach to Einstein-Hilbert type actions 
using structures of generalized geometry 
can be found in Ref.~\refcite{JV}.

%\subsection{Subsections only have the first letter of the entire title capitalized}

%Subsections only have the first letter of the first word capitalized (except for words that are naturally capitalized).

\section{Background setup in generalized geometry}

We consider a vector bundle $E = TM \bigoplus T^*M$.
The elements of the space of sections of the vector bundle are formal sums $e = X + \zeta \in \Gamma(E)$, where $X$ is a vector field and $\zeta$ is a one-form.
We have a natural pairing 
\begin{equation}
\langle X+\zeta , Y + \eta \rangle = i_X \eta + i_Y \zeta \, ,
\label{pairing}
\end{equation}
which is symmetric and non-degenerate.
The signature of the pairing is $(d,d)$, where $d$ are the spacetime dimension of $TM$ and $T^*M$ respectively. 
The pairing is invariant under $O(d,d)$ transformations.
We have also a Dorfman bracket
\begin{equation}
[X+\zeta , Y+\eta]_{\text{D}} =[X,Y]_{\text{Lie}} + \mathcal{L}_X \eta - i_Y d\zeta \, ,
\end{equation}
where $[X,Y]_{\text{Lie}} $ is the Lie bracket of vector fields.
Finally there is an anchor map
$a: E \rightarrow TM$ 
that maps from the vector bundle being considered here to the tangent bundle.
Thus we define a Courant algebroid:
$(E, \langle \; , \; \rangle, [\; ,\;]_{\text{D}}, a)$,
with the following properties:
\\
for a function $f \in C^{\infty} (M)$ and elements $e_1, e_2 \in \Gamma(E)$, 
%$-$
the Dorfman bracket $[\; , \;]_{\text{D}} : \Gamma(E) \times \Gamma(E) \rightarrow \Gamma(E)$ satisfies the Leibniz rule
\begin{equation}
 [e_1 , f e_2]_{\text{D}} = f[e_1,e_2]_{\text{D}} + (a (e_1) f) e_2 
 \label{Leib}
\end{equation} 
and Jacobi identity
\begin{equation}
 [e_1,[e_2,e_3]_{\text{D}}]_{\text{D}} = [[e_1,e_2]_{\text{D}}, e_3]_{\text{D}} + [e_2, [e_1, e_3]_{\text{D}}]_{\text{D}} \, .
 \label{Jacobi}
\end{equation}
%$-$
The anchor map $a$ obeys the homomorphism property 
\begin{equation}
a \left( [e_1 , e_2]_{\text{D}} \right) = \left[a(e_1) , a(e_2) \right]_{\text{Lie}} \, , 
\end{equation} 
while the pairing 
 $ \langle \; , \; \rangle: \Gamma(E) \times \Gamma(E) \rightarrow C^{\infty}(M)$ 
exhibits the following properties, 
\begin{equation}
a (e_1) \langle e_2 , e_3 \rangle = \langle [e_1,e_2]_{\text{D}}, e_3 \rangle + \langle e_2 , [e_1,e_3]_{\text{D}} \rangle
\label{killing}
\end{equation}
and
\begin{equation}
a^{\dagger} d \langle e_1,e_2 \rangle = 
[e_1, e_2]_{\text{D}} + [e_2,e_1]_{\text{D}} \, ,
\label{notLie}
\end{equation}
where $a^{\dagger}: T^*M \rightarrow E^* \simeq E$.
From (\ref{notLie}), it is obvious that the Dorfman bracket is not a Lie bracket as it is not anti-symmetric. 
\\
\\
{The following are examples of $O(d,d)$ transformations.}
\\
{$B$}-transform: 
\begin{equation}
e^{B} 
% \begin{pmatrix} V \\ \zeta \end{pmatrix} 
{V \choose \zeta}
= { 1 \quad 0 \choose B^T \quad 1} {V \choose \zeta}
\end{equation}
%\flushleft
\begin{equation}
B: TM \rightarrow T^*M  \quad \text{where} \quad B \in \Omega^2 (M)
\; \text{is a two-form}.
\nonumber
\end{equation}
%\noindent
This orthogonal transformation is well known in string theory and will be a central object in our current study.
\\
$\beta$-transform:
\begin{equation}
e^{\beta}
{V \choose \zeta}
= { 1 \quad \beta^T \choose 0 \quad 1} {V \choose \zeta}
\end{equation}
{\flushleft
\begin{equation}
\beta:  T^*M \rightarrow TM  \quad \text{where} \quad \beta \in \mathfrak{X}^2 (M) \; \text{is a bi-vector}.
\nonumber
\end{equation}}
We refer interested readers to Ref.~\refcite{JurcoSchupp} for an application in non-commutative geometry.
\\
{Diffeomorphism:}
\begin{equation}
O_N 
% \begin{pmatrix} V \\ \zeta \end{pmatrix} 
{V \choose \zeta}
= \begin{pmatrix}
N^T & 0\\
0 & N^{-1}\\
\end{pmatrix} {V \choose \zeta}
\end{equation}
{\flushleft
\begin{equation}
N: TM \rightarrow TM \quad \text{where} \quad  N|_{p \in M} \in GL(d).
\nonumber
\end{equation}}

\section{Deformations}

%Use the ``graphicx" package following the example below  to include .EPS figure files. Word template files are discouraged, allowed as a last resort for those people who have some difficulties with \LaTeX.

%Since the World Scientific proceedings style \cite{ws} uses numbered superscript citations of the bibliography items, one has to be careful to use 
%\verb|\refcite| to get a baseline normal size number to include in an in-line direct reference, best formatted in the style: 
 %``see Ref.$\sim$\verb|\refcite|\{\ldots\}" while normal superscript citations follow punctuation  as in ``.\verb|\cite|\{\ldots\}"  For example, here is an in line citation to Ref.~\refcite{arxiv}. On the other hand citations at the end of a sentence are done line this.\cite{mg14}

%Consult the full World Scientific instruction file for details if problems arise with your normal \LaTeX\ writing style.

Within the context of satisfying the Courant algebroid properties (\ref{Leib}) - (\ref{notLie}),
we propose the following deformations 
\begin{equation}
\langle e_1 , e_2 \rangle \rightarrow 
\langle e_1 , e_2 \rangle ^{'} =
\langle e^{\mathcal{G}} (e_1) , e^{\mathcal{G}} (e_2) \rangle
\label{defi}
\end{equation}
\begin{equation}
[e_1 , e_2]_{\text{D}} \rightarrow 
[e_1 , e_2]_{\text{D}}^{'} =
e^{-\mathcal{G}} [e^{\mathcal{G}} (e_1) , e^{\mathcal{G}} (e_2)]_{\text{D}} \, ,
\label{defii}
\end{equation}
for elements $e_1 = X + \zeta$, $e_2 = Y + \eta$.
We have introduced here a non-symmetric metric { $\mathcal{G} = g + B$}, 
which is composed of a symmetric $g$ and an anti-symmetric $B$ as an 
invertible map
$\mathcal{G}: TM \rightarrow T^*M$
and
$e^{\mathcal{G}}: E \rightarrow E$
: $e^{\mathcal{G}} (e) = e + \mathcal{G}\left(a(e),-\right)$.

\section{Computations}
The deformations (\ref{defi}) and (\ref{defii}) preserve the Courant algebroid properties. \\
Given elements $X + \zeta$ and $Y + \eta$,
deformation (\ref{defi}) corresponds to 
the pairing being deformed with the symmetric metric $g$,
\begin{equation}
\langle X + \zeta , Y + \eta \rangle ^{'} =
\langle X + \zeta , Y + \eta \rangle + 2g \left(X , Y \right) \, , 
\end{equation} 
while for the deformed Dorfman bracket, 
it is straightforward to compute
from (\ref{defii}) by using the definition of Dorfman bracket 
that
\begin{eqnarray}
&& [X + \zeta , Y + \eta]_{\text{D}}^{'} 
\nonumber
\\
 =
 &&
  [X + \zeta , Y + \eta]_{\text{D}} +
X\mathcal{G}(Y,-) - Y \mathcal{G} (X,-) +  d\mathcal{G} (X,Y) 
\nonumber
\\
&& 
- \mathcal{G} (Y,[X,-]_{\text{Lie}}) - \mathcal{G}  ([X,Y]_{\text{Lie}},-) + \mathcal{G} (X,[Y,-]_{\text{Lie}}) 
\nonumber 
\\
=
&&
 [X + \zeta , Y + \eta]_{\text{D}} + 2 g
 (\nabla X, Y ) \, .
\label{newD}
\end{eqnarray}
We find that the bracket is twisted by a connection $\nabla$ in which the non-symmetric metric $\mathcal{G} = g+B$ is encoded.
From (\ref{newD}),
\begin{eqnarray}
2g(\nabla_Z X, Y ) 
=
&&
X\mathcal{G}(Y,Z) - Y \mathcal{G} (X,Z) + Z \mathcal{G} (X,Y) 
\nonumber
\\
&& 
- \mathcal{G} (Y,[X,Z]_{\text{Lie}}) - \mathcal{G} ( [X,Y]_{\text{Lie}},Z) + \mathcal{G} (X,[Y,Z]_{\text{Lie}}) 
\label{genKos}
\end{eqnarray}
is a generalized version of the Koszul formula that involves torsion, while the original formula defines the torsion-free Levi-Civita connection.
Note the unusual ordering of arguments in (\ref{genKos}).
From the generalized Koszul formula (\ref{genKos}), we compute the torsion connection
\begin{equation}
g (\nabla_X Y,Z) = g(\nabla^{LC}_X Y,Z) + \frac{1}{2} H(X,Y,Z) \, ,
\label{gencon}
\end{equation}
where $H (= dB)$ is an anti-symmetric 3-form and $\nabla^{LC}$ is the Levi-Civita connection. \\
For contortion  
\begin{equation}
K(X,Y,Z) = 
\frac{1}{2} \,
\left( g(T(X,Y),Z ) + g(T(Z,X), Y) + g (T(Z,Y), X) \right) \, ,
\end{equation}
%From deformed (\ref{killing}), we get
%\begin{equation}
%2g(\nabla_X Y,Z) + 2g(\nabla_Z Y,X) 
%= 2 [
%Yg(Z,X) - g([Y,Z]_L , X) - g(Z, [Y,X]_L)
%]
%\end{equation}
we can deduce from the deformed property (\ref{killing}) that
\begin{equation}
2K(X,Y,Z) = H(X,Y,Z) = g(T(X,Y), Z) \, . 
\end{equation}
On the other hand, from the deformed version of property (\ref{notLie}), we have metricity
of the connection 
\begin{equation}
g (\nabla_X Y,Z) + 
g (\nabla_X Z,Y)  
= Xg(Y,Z) \, .
\end{equation}

\section{Results}

The connection that appears in the deformation is found to be
\begin{equation}
g \circ \nabla = g \circ \nabla^{LC} + K 
\label{conn}
\end{equation}
with contortion $K$.
We find that the correspondingly deformed equation
(\ref{killing})
gives us a totally anti-symmetric contortion $K = H/2 = dB/2$.
The contortion is closed under the deformed Jacobi identity (\ref{Jacobi}),
whereas the deformed equation
(\ref{notLie})
gives us the metricity condition, 
see Ref.~\refcite{Jan}
for further computational details and results.
\\
\\
Having the connection (\ref{conn}), we compute the
Ricci tensor in components
\begin{equation}
R_{jl} = R^{LC}_{jl} - \frac{1}{2} \nabla^{LC}_i H_{jl}^{\phantom{jl}i} 
- \frac{1}{4} \, H_{lm}^{\phantom{lm}i} H_{ij}^{\phantom{ij}m} \, .
\label{Ric}
\end{equation}
It turns out to be non-symmetric due to the anti-symmetric second term.
When (\ref{Ric}) is treated as a vacuum field equation, that is, let
$ R_{jl} = 0$,
we have the corresponding non-symmetric gravity action 
\begin{equation}
S_{\mathcal G} = \frac{1}{16\pi G_{N}} \int d^{d} x \sqrt{-g} 
\left( R^{LC} - \frac{1}{12} H_{ijk}H^{ijk}  \right) \, ,
\label{nmg}
\end{equation}
where $G_N$ is Newton's gravitational constant in $d$ dimensions.
Varying with respect to $g$ and $B$ implies the field equations:
 \begin{equation} 
\frac{\delta S}{\delta g^{lj}} = 0 \, , \quad   
\frac{\delta S}{\delta B^{lj}} = 0 
\qquad
\Rightarrow 
\qquad
R_{jl} = 0
\, .
\end{equation}
Note the ordering of the indices, which is  
only a matter of convention.

\section{Discussions}
In string theory, the non-linear sigma model on worldsheet $\Sigma$,
with worldsheet metric $ \gamma^{\mu\nu}$ for $\mu, \nu = 0,1$,
\begin{eqnarray}
S_{\text{nlsm}} =&& \frac{1}{4\pi\alpha'} \int_{\Sigma} d^2 \sigma \sqrt{\gamma}    
\;  
\nonumber
\\&&
(\gamma^{\mu\nu} {g_{mn} (X)} \, \partial_{\mu} X^{m} \partial_{\nu} X^{n} 
\nonumber
\\&& \, + i\epsilon^{\mu\nu} {B_{mn} (X)} \, \partial_{\mu} X^{m} \partial_{\nu} X^{n}
\nonumber
\\&& \, + \alpha' \phi(X) R_{(\gamma)})
\, ,
\label{nlsm}
\end{eqnarray}  
where $m,n = 0, 1, \cdots, 25$, in $26$-dimensional spacetime,
describes the string propagation in background fields: metric $g$, Kalb-Ramond $B$ and dilaton $\phi$. 
Beta functions
\begin{eqnarray}
\beta_{\mu\nu}(g) &=& \alpha' R^{LC}_{\mu\nu} - \frac{\alpha'}{4}
H_{\mu\lambda\kappa}
H_{\nu}^{\phantom{\nu}\lambda\kappa}
+ 2 \alpha'\nabla_{\mu}\nabla_{\nu}\phi
\label{b1}
\\ 
\beta_{\mu\nu}(B) &=& -\frac{\alpha'}{2}\nabla^{\lambda} H_{\lambda\mu\nu} 
+ \alpha'\nabla^{\lambda}\phi H_{\lambda\mu\nu}
\label{b2}
\\
\beta_{\mu\nu}(\phi) &=& -\frac{\alpha'}{2}\nabla^2 \phi +
\alpha' \nabla_{\mu}\phi\nabla^{\mu}\phi
- \frac{\alpha'}{24} H_{\mu\nu\lambda} H^{\mu\nu\lambda} \, ,
\label{b3}
\end{eqnarray}
which follow from (\ref{nlsm}) are required to vanish in order to  
preserve the Weyl invariance in string theory as a quantum theory.
The low-energy closed bosonic string action (Ref.~\refcite{Callan})
\begin{equation}
S_{\text{eff}} = \frac{1}{2\kappa^2} \int d^{26} X \sqrt{-g} \; e^{-2\phi} 
\left( R^{LC} - \frac{1}{12} H_{abc}H^{abc} + 4 g^{ab}\partial_a\phi \, \partial_b\phi \right) 
\label{eff}
\end{equation}
has been derived as the effective string action that gives 
the equations of motion, which are equal to 
the vanishing beta functions (\ref{b1}), (\ref{b2}) and (\ref{b3}).  
\\
\\
We notice that our non-symmetric Ricci tensor (\ref{Ric}) 
contains the beta functions (\ref{b1}) and (\ref{b2}) and our non-symmetric gravity action (\ref{nmg}) resembles the closed string effective action (\ref{eff}) without dilaton. 
Our action (\ref{nmg}) is effectively an action,
where its equations of motion describe a Ricci flow.  
A closely related work has previously appeared in
Ref.~\refcite{GF} in the scope of supergravity.
\\
While in string theory, the spacetime dimension of (\ref{eff}) is determined 
during the derivation of the beta function (\ref{b3}),
the dimension $d$ of our similarly Ricci-flat spacetime theory (\ref{nmg}) is unrestricted.
Recall that our Courant algebroid deformations involve only the tangent bundle.
Interestingly, 
the combination of $g + B$ in the non-linear sigma model (\ref{nlsm}), 
which was a motivation in the pursuit of non-symmetric gravity theory,
appears to be at equal footing in our deformations (\ref{defi}) and (\ref{defii}).
Deforming the Einstein-Hilbert action has led us to an Einstein-Kalb-Ramond theory 
(Ref.~\refcite{SG}).

\section*{Acknowledgements}
B. Jur\v co acknowledges the
GA\v CR P201/12/G028 grant and hospitality of MPIM.
F.\,S. Khoo, P. Schupp and J. Vysok\' y would like to thank DFG RTG-Models of Gravity for grant support.

\end{document}